\documentclass[a4paper,UKenglish]{lipics}
 
\usepackage{microtype}
\title{Teaching Logic for Computer Science: \newline Are We Teaching the Wrong Narrative?
}
\titlerunning{Teaching Logic for CS} 

\author[]{Johann A. Makowsky
\footnote{President of EACSL (European Association of Computer Science Logic) 2004-2009}
}
\affil[]{
Faculty of Computer Science \\
Technion--Israel Institute of Technology, 
Haifa Israel \\
\texttt{janos@cs.technion.ac.il}}
\authorrunning{J.A. Makowsky} 

\Copyright{Johann A. Makowsky}

\subjclass{K 3.2 Computer and Information Science Education}

\serieslogo{logo_ttl}
\volumeinfo
  {M. Antonia {Huertas}, Jo\~ao {Marcos}, Mar\'ia {Manzano}, Sophie {Pinchinat}, \\
  Fran\c{c}ois {Schwarzentruber}}
  {5}
  {4th International Conference on Tools for Teaching Logic}
  {1}
  {1}
  {101}
\EventShortName{TTL2015}
\usepackage{calc}
\usepackage{ifthen}
\usepackage{epsfig}
\usepackage{graphicx}
\usepackage{latexsym}
\usepackage{amssymb,amsmath}

\newif\ifshort
\shorttrue
\newif\ifskip
\skiptrue

\renewcommand{\theenumi}{\roman{enumi}}

\newcommand{\SOL}{\mathrm{SOL}}
\newcommand{\FOL}{\mathrm{FOL}}

\newcommand{\N}{{\mathbb N}}

\begin{document}
\maketitle
\begin{abstract}
In this paper I discuss what, according to my long experience, every
computer scientist should know from logic. We concentrate on issues of modeling,
interpretability and levels of abstraction. We discuss what the minimal toolbox of
logic tools should look like for a computer scientist who is involved in designing
and analyzing reliable systems. We shall conclude that many classical topics dear to
logicians are less important than usually presented, and that less-known ideas from
logic may be more useful for the working computer scientist.
\end{abstract}
\sloppy
\section{Introduction}
{\em 
\footnotesize
Even in teaching mathematics we can at least attempt to teach the students the flavor of 
freedom and critical thought, and to get them used to the idea of being treated as humans 
empowered with the ability of understanding. 
\hfill  \rm Roger Godement, Cours d'Alg\`ebre, Hermann, Paris 1966} 
\vspace{0.5cm}

In the last few years we see contradictory developments concerning teaching logic
for Computer Science undergraduates. On the one side, the importance of logical tools
in hardware and software engineering, artificial intelligence, and database management, including the
new trend of handling large data, is widely recognized. The Vienna Summer of Logic of 2014\footnote{
Vienna Summer of Logic 2014 http://vsl2014.at/
} attracted over 2500 researchers, but the traditional central logic conferences (ASL Logic Colloquium, LICS and CSL)
played 
only minor part in terms of attendance. 
Most attendees were practitioners of logic, benefiting from and contributing to the
the {\em unusual effectiveness of logic in computer science}, \cite{halpern2001-unusual}.
On the other side, leading Computer Science Departments\footnote{
ETHZ Zurich, is a good example. There was strong support to keep logic in the curriculum, but finally it was dropped,
after the time and priorities argument won the vote by a small margin}
have taken logic courses off their compulsory curriculum
of the 3 year program. The argument behind such  a decision does not question the importance of logic,
but cites the scarcity of available time for teaching the essentials.

The discrepancy between the success of logic in Computer Science and its relegation from the Computer Science curricula
may indicate that the standard logic courses are outdated and miss their point.
Teachers of logic in Computer Science are  often still teaching courses which are a mix of formalizing logical reasoning, meta-mathematics, 
and the fading reverberations of the famous crisis of the foundations of mathematics.
By doing so, they are contributing to the disappearance of their courses from mainstream  undergraduate education in
Computer Science.
We have to rethink which aspects of logic matter for the Computer Science undergraduate programs.
I want to argue, that the fascination with completeness and incompleteness theorems, the
beauty of the compactness theorem  and its applications in infinite model theory, 
and the focus on first order logic in general, are didactically counterproductive.

I was trained as a mathematician with a PhD  in logic (1974, ETHZ Zurich), in model theory, to be precise.
I started publishing in Computer Science in 1980, in database theory, logic programming and finite model theory.
For the last twenty years I mostly worked on applications of logic to combinatorics \cite{bk:GroheMakowsky2011}.
I have been teaching logic for Computer Science and other logic related courses 
for over thirty years at the Technion's Computer Science Department. I have designed the current
version of the compulsory logic course, called 
{\em Sets and   Logic for Computer Science}\footnote{http://webcourse.cs.technion.ac.il/234293/}.
It consists of three hours per week frontal lectures and two hours per week tutorials in a thirteen week
semester.  It is given twice or even three times a year with close to 400 students enrolling each year.
There are various teachers teaching the course,  some of them leading researchers who use logic in their work,
some of them 
colleagues who 
are well educated in logic and
happen to like teaching logic, without logic playing a major part in their
research. This makes introducing changes in the course more difficult.
I tried and still try to shift the emphasis of this course. 
I lectured about the rational behind this course and the changes I want to introduce
at various conferences, formally at LPAR 2007, CiE 2008
\cite{conf-lpar-Makowsky07,makowsky2008a-hilbert},
and informally at Computer Science in Russia, Moscow 2008.
I lectured about it also at ETH Zurich and the Technical University in Vienna.
A journal version appeared as \cite{makowsky2008-hilbert}.

I have received many comments, criticisms, and encouragement\footnote{
I would like to thank Nadia Labai for her critical reading of this paper,
and to
A. Avron, 
K. Censor-Hillel,
M. Kaminski, and two anonymous referees for valuable comments and suggestions.
}.
In this paper I want to
sharpen my point of view  by addressing many of the critical comments I have received in the past.

\section{Learning from the Past: Linear Algebra}
We first look at the introduction of Linear Algebra
into the curriculum of Mathematics and Physics students in the 1950ies. 
It took more than ten years to become a mainstream policy to teach Linear Algebra in the first year.
We shall see in the sequel that logic courses in undergraduate Computer Science
fail to achieve similar goals because they focus on the wrong issues.

The arguments for introducing Linear Algebra were twofold. Students should be exposed to
Bourbaki-style abstract thinking and they should learn the tools needed
later in courses like Numeric Analysis, Differential Equations, Functional Analysis, 
Mathematical Physics,  and Quantum Physics. Linear Algebra seemed ideal for both purposes.
On the abstract side, 
linear functions between vector spaces are distinguished from 
their representations via matrices over the underlying field. 
This is a first example
where representation of (matrix-syntax) and being a 
linear function (function-semantics) is distinguished.
Solving linear equations is a semantic problem, whereas computing a determinant is a syntactic method.
The proposed course would have two goals: normal forms of matrix representations and
the foundations of the determinant method. 
One  really proves a {\em completeness theorem}:
A system of linear equations has a 
unique 
solution (semantics) iff the determinant representing these
equations does not vanish (syntax).
The introduction of such a course was a big success for two reasons:
It provided the student with a theoretical framework to be used
over and over again. More importantly however, the concepts lived later on and were 
applicable in the most diverse situations.
I have emphasized here the syntax/semantics dichotomy. The truth, however, is that most teachers of such a course
are not aware of this distinction.

For some strange reasons, versions of Linear Algebra or Modern Algebra\footnote{The original title of van der Waerden
landmark monograph \cite{waer:48}, where  the word ``modern'' was dropped long ago.}
are still taught almost the same way in the first year of a Computer Science curriculum. 
The reasons are strange, because it is  not only conceivable but entirely
normal that students of Computer Science will never encounter determinants in their entire three year program.
If a student nevertheless encounters these concepts later on, be it in Numeric Analysis, Coding Theory or Complexity,
the teacher feels obliged to develop the concepts from scratch.
In other words such a course given in the first year is a total waste of time.

\section{Theoretical Orientation vs Practical Knowledge}
It is useful to distinguish between {\em theoretical orientation}
and {\em practical knowledge}.
Most electricians 
do have a lot of practical knowledge
which they can apply when installing or repairing wiring and appliances.
They may have a vague knowledge  of the physics and electrodynamics on which their practical
knowledge is based, but they do not have to understand the Maxwell equations.
Their theoretical orientation is very limited.
Structural engineers must have a very sophisticated practical knowledge of material science
and applied mathematics, but again their theoretical orientation concerning the foundations of
physics and real and complex analysis remains vague.

Our example of the Linear Algebra course is more balanced in this respect. It tries to convey
the level of abstraction needed to understand (rather than use) the tools of matrix calculus,
and it does also teach how to use those tools. In Linear Algebra courses for engineers the roles
of understanding vs using may be shifted in favor of using.

So what does this mean for teaching Sets and Logic for Computer Science?
\section{The Traditional Narrative}
When logic and set theory courses were introduced in the Mathematics curriculum at about the same time
as linear algebra the main purpose was theoretical orientation. The students should be
exposed to the proposed solution of the so called crisis in the foundations of mathematics.
The standard textbooks were \cite{bk:Halmos} by P. Halmos for naive set theory and
the books  \cite{mendelson1997introduction,lyndon1966notes,enderton2001mathematical} by E. Mendelson, R.C. Lyndon and
H. Enderton respectively, for mathematical logic. Halmos' book was reprinted in 1974, but now K. Devlin's \cite{devlin1993joy}
has taken its place.  The books by Mendelson and Enderton are still reprinted and used.
Lyndon's text teaches logic, provided the reader already knows abstract algebra.
More modern textbooks I like are \cite{bk:EFT94,bk:EF95}\footnote{
For mathematicians there are also  the classics \cite{bk:shoe,bk:Monk76,manin2010course}. 
My true favorite is \cite{adamowicz2011logic} by S. Adamowics and P. Zbierski.}.

Naive set theory teaches the student to use set theory for defining ordinals and cardinals and their arithmetic.
Sometimes Russel's' paradox is discussed as making a problem, sometimes as a hint that not all concepts
can be formalized as sets.
The axiom of choice and well-orderings play a prominent role. Very little of this, however, is relevant for our
Computer Science undergraduates. I will briefly sketch what may be relevant for them in the next section.

The logic texts focus on First Order logic $\FOL$. They define first order formulas and their interpretations.
They give the semantic notions of logical consequence and logical equivalence. Then they give  the syntactic
notions of deduction rules and proofs (proof sequences). They may distinguish between natural deduction (Gentzen style)
and Hilbert style deductions. They prove the completeness and compactness theorems, the L\"owenheim-Skolem-Tarski theorems,
and show that neither the real number field nor the arithmetic structure of the natural numbers
can be characterized in First Order Logic. They may prove G\"odel's incompleteness theorems. Again,
this may give some theoretical orientation. 
But none of this is the raison d'\^etre why 
our Computer Science undergraduates should learn logic. 
There is no practical knowledge gained from proving the Completeness and Compactness Theorems in full rigor. 

The first proper modern logic monograph was \cite{bk:HilbertAckermann-3ed} by D. Hilbert and W. Ackermann, 
which appeared first in 1928, and in English as \cite{bk:HilbertAckermann-eng} in 1950.
One should emphasize that for Hilbert and Ackermann, logic was Second Order Logic,  
and what we call First Order Logic
is called in their book the {\em restricted calculus}. They gave an axiomatization of the restricted calculus
and asked whether their axiomatization is complete. K. G\"odel, reading the book in its first year of
publication, showed immediately that the answer was positive. As it follows from G\"odel's incompleteness theorem
that Second Order Logic has no recursive axiomatization, Second Order Logic was deemed to be too much like set theory,
and First Order Logic took center stage. Still, the natural language to describe mathematics is second or even higher order 
logic\footnote{
I still was a witness  in 1966 of an argument between P. Cohen and P. Bernays on whether Cohen had proved the independence
of the Continuum Hypothesis from set theory. Bernays insisted that Hilbert's problem was not about
first order provability  in Zermelo-Fr\"ankel set theory.}.

\section{For Whom Should We Teach Sets and Logic in Computer Science?}
We take our clue from the discussion about Linear Algebra.
What we teach in Sets and Logic should be visibly used in the following courses:

\ifskip
\begin{itemize}
\item
Data Structures and Algorithms
\item
Formal Languages, Automata Theory, and Computability
\item
Database Systems
\item
Graph Algorithms and Complexity
\item
Formal Methods and
Verification, in all their ramifications
\item
Decision Procedures in Automated Theorem Proving
\end{itemize}
\else
\fi

If the undergraduate curriculum contains in one form or another at least three of the above topics,
a course like Sets and Logic should be taught. What remains to be discussed is the choice of topics
and their emphasis.

\section{The World of Sets}
Our undergraduates do not need an introduction into cardinal arithmetic, but they do need an understanding 
and proficiency in handling statements like

\begin{itemize}
\item
An ordered pair $\langle a , b \rangle$ is a set with the {\em basic property}
$\langle a , b \rangle  = \langle a' , b' \rangle$  iff $a=a'$ and $b=b'$.
\item There are various realizations of ordered pairs:
\begin{enumerate}
\item
$\langle a , b \rangle_{Ku1} =\{ \{a\}, \{a,b\} \}$.
\item
$\langle a , b \rangle_{Ku2} =\{ a, \{a,b\} \}$.
\item
$\langle a , b \rangle_{Wie} = \{ \{ \{ a \} , \emptyset \}, \{\{ b \}\}\}$.
\end{enumerate}
\item
A finite automaton is a  quintuple $\langle S, \Sigma, \delta, s, F \rangle$
where $S$ is a finite set (of states), $\Sigma$ is a finite set of symbols (an alphabet), 
$\delta: S \times \Sigma \rightarrow S$ is a (transition) function, $s \in S$ is a starting state, and $F \subseteq S$
is the subset of final (accepting) states.
\end{itemize} 

Additionally, they do need a repertoire of set constructions\footnote{
Basically these are the constructions needed to build the cumulative hierarchy}.
which allows them to construct sets which are realizations
of the concepts we use in Computer Science: 

\begin{itemize}
\item
Sets, relations, functions, Cartesian products, infinite unions;
\item
Finite sets, countable sets, uncountable sets, equipotent sets;
\item
The set of words $\Sigma^{\star}$ over an alphabet $\Sigma$ and the set of natural numbers $\N$.
\item
Inductively defined sets,  such as  the well-formed formulas in some formal language, etc.
\end{itemize}

Some of this material is usually covered in the beginning of high-level textbooks such as \cite{bk:HMU}.
To the usual material, I would add without proof and discussion of the axiom of choice, also the statements: 

\begin{itemize}
\item
The Cartesian product of a non-empty (finite) 
family of non-empty sets is not empty.
\item
The union of countably many countable sets is countable.
\end{itemize}

\ifskip
However, I would advise not to talk about the 
axiom of choice or the well-ordering principle, and cardinalities of sets.
For our purpose it suffices to talk about equipotent sets.
Sets are finite, countable, or equipotent to some previously constructed set, 
such as the powerset of the natural numbers,
or, equipotent to it, the set of functions $f: \N \rightarrow \{0,1\}$.
\fi

\section{Side Effects}
One aspect is never covered in the usual introduction to the world of sets:
The realization of concepts as sets has {\bf side effects}.

Side effects are properties of the realization which are not intended.
In the example of the three realizations of ordered pairs, $a$ is an element only of $\langle a , b \rangle_{Ku2}$,
$b$ is not an element of any of them, $\{a\}$ is an element of $\langle a , b \rangle_{Ku1}$ only,
and $\{\{b\}\}$ is an element of $\langle a , b \rangle_{Wie}$ only.
The further difference between $\langle a , b \rangle_{Ku1}$ and $\langle a , b \rangle_{Ku2}$
is that to prove the basic property for 
$\langle a , b \rangle_{Ku1}$,
extensionality suffices, whereas for 
$\langle a , b \rangle_{Ku2}$
the axiom of foundations is needed. 
This observation is needed to explain why $\langle a , b \rangle_{Ku1}$ is preferable over
$\langle a , b \rangle_{Ku2}$. It is also preferable over
$\langle a , b \rangle_{Wie}$, because  $\langle a , b \rangle_{Wie}$ uses the empty set as part of its definition,
whereas the other two versions use only curly brackets and the sets $a$ and $b$.

We can look also at two definitions of natural numbers, both starting with the empty set realizing the
element $0$. For $\N_{naive}$ we define the successor of $n$ as $\{n\}$. For $\N_{vNeu}$ we define
the successor of $n$ as $n \cup \{n\}$. 
They both satisfy the Peano Postulates. The side effects are that
in the first every $n \neq 0$ has exactly one element, whereas in the second every $n$ is equipotent
to the set of its predecessors.

The notion of side effects, as properties of a realization or implementation which are not intended,
is central to understanding hardware and software systems. It can be easily explained in the world of sets,
and should be a {\em leitmotiv} throughout 
all the courses taught.

\section{Syntax and Semantics}
Having learned to master inductive definitions we should now introduce the syntax of logical formalisms.
This looks like a  further exercise in inductive definitions.
We define the formalisms of Quantified Propositional Logic and prove the Unique Readability Theorem.
The meaning of a formula is given by the {\em meaning function} which associates with a formula its
meaning: In the case of (Quantified) Propositional Logic this is a Boolean function. 
We can introduce the notions of logical equivalence
and consequence, and prove normal form theorems and quantifier elimination.
We can introduce proof sequences and state, but not prove, the completeness theorem for the proof
sequences introduced.  Proving it requires to much time which we need for other purposes.

As said before,  Second Order Logic $\SOL$ is the logic which allows us 
to talk about most important concepts, say in graph theory or formal language theory.
Therefore we introduce the syntax of $\SOL$ right away. 
We start with a vocabulary (set of relation and function symbols
including constant functions which are constants). The meaning (interpretation) of a vocabulary $\tau$
is  a $\tau$-structure. 
The meaning (given by the meaning function) of an $\SOL(\tau)$-formula with free first order variables 
is a {\em relation in a $\tau$-structure}.
If there are no free variables, it is a Boolean value, 
but it is preferable to treat this as a special case and keep the case with
free variables in the center of our attention. 
If there are free second order variables in the formula, the meaning is given by sets of relations.

\section{Read and Write}
Before we embark on proving meta-theorems the students should have some minimal
proficiency in reading and writing $\SOL$-formulas.
The mathematics students learning logic can be assumed to be familiar
with mathematical statements from solving equations and from analytic geometry.
For them, the notion of the geometric locus of all points satisfying a set of equations
might come natural.
The Computer Science students have no such background.
I am always perplexed to see how difficult it is, even for
advanced Computer Science students, to correctly  write down an $\SOL$-formula
expressing that a graph is connected, or Hamiltonian. The difficulty is
real (and mathematical) when trying to formulate that a graph is planar, because we have to use
Kuratowski's or Wagner's Theorem. 
Once the students have written down a formula which supposedly expresses what they had in mind,
it remains difficult for the students to read their own formula in order to check its correctness.

Practicing reading and writing skills by drawing graphs and finding $\FOL$ formulas which
distinguishes the graphs, or internalizing that isomorphic structures cannot be distinguished,
is a prerequisite for all further developments. This brings us back to side effects.
It is also a side effect that two isomorphic structures are different from the point of view
of their definition in the language of sets.

\section{Pebble Games Help US Understand Quantification}
Students encounter great difficulties in grasping the order of quantification even in $\FOL$.
It might be useful to introduce pebble games at an early stage. 
They are very visual and leave an impression on the students.
Reading off the winning strategy from formulas in prenex normal form is very visual and easy to understand
and to prove.
So if a formula with $n$ quantifiers distinguishes two structures, player I has a winning strategy.
The converse, if no formula with $n$ quantifiers distinguishes two structures, then player II has winning strategy,
can be stated without proof. 
The easy part can be used to show that

\begin{itemize}
\item
One needs at least $n$ existential quantifiers to say that there are at least $n$ elements in 
the structure over the empty vocabulary.
\item
One cannot express that there is an even number of elements in the structure over the empty vocabulary.
\end{itemize}

If we look at a the graph of the gender-free parent relation, there are natural concepts like grandparents,
siblings, cousins, and even third-cousins twice removed, which can be expressed on $\FOL$.
Other concepts, like being an ancestor, or the requirement that the parent relation be cycle-free,
are not expressible in $\FOL$, even if restricted to finite structures.
Again, to show the non-definability in $\FOL$ one can use the pebble games.

One should also prove that there are $\forall \exists$-formulas in $\FOL$ which are not equivalent to
any $\exists\forall$-formulas, and vice versa. Again this is not difficult using 
the properties  that universal formulas are preserved under substructures (which can be modified to
$\exists\forall$-formulas), and that $\forall \exists$-formulas are preserved under unions of chains.

All this is more relevant for the later courses than proving the completeness or compactness theorem.

\section{Definability and Non-definability}

Definability is a central concept of Logic.
We can define concepts in the language of sets using the membership relation only.
These are the set-definable concepts. The realization of these concepts as sets have  many properties expressible in the language of sets
from which we want  to abstract.
By defining concepts as $\tau$-structures we fix the level of abstraction,
and distinguish between the essentials and the side effects.

Every $\SOL$-definable concept is also set-definable,
but the converse is not true. Topological spaces are good examples for mathematicians.
They are not $\tau$-structures in the sense we have in mind.
For our Computer Science undergraduates one can explain the difference between set-definability and
$\SOL$-definability by anticipating the notion of computable
languages (sets of words): Every computable language is set-definable, but $\SOL$-definable languages have
bounded complexity. 
$\FOL$-definable concepts are computationally even simpler. 
Logical formalisms are used to compute definable concepts. We choose the formalisms
to be expressive enough for use in a particular application, but to be computationally feasible enough
to allow implementation.

\section{Conclusions}

We have suggested that the meta-mathematical narrative of the  traditional logic courses
for Computer Science is missing its purpose and contributes to the disappearance of logic
from the undergraduate curriculum.

We have emphasized that the world of sets is used for modeling concepts of Computer Science
using simple set constructions and inductive definitions. It is also used to prove properties
of these concepts.  We suggested to introduce, first as an example of long inductive definitions,
quantified propositional logic and its meaning, given by Boolean functions.
This is used to develop the basic concepts of logic, logical equivalence and consequence.
We suggested as a next step to introduce Second Order Logic, and leave First Order Logic as
a special case. We put the emphasis on teaching how to read and write properties of the modeled concepts,
and to concentrate on the expressive power of Second Order Logic and its fragments.
Finally we suggested to concentrate on tools to prove non-definability such as pebble games and
preservation properties.

So where are all the beautiful theorems to be taught? In follow up courses when needed.
Completeness and incompleteness in a course on decision procedures of logical systems,
or in a course on proof theory, compactness in a course on model theory for mathematicians, 
or in a course on the foundations of logic programming. None of these will be undergraduate courses.
In contrast to this, what we suggest to teach is used in

\begin{itemize}
\item
Data Structures and Algorithms
\item
Formal Languages, Automata Theory and Computability
\item
Database Systems
\item
Graph Algorithms and Complexity
\item
Formal Methods and
Verification, in all their ramifications
\item
Decision Procedures in Automated Theorem Proving
\end{itemize}

Explaining Completeness and Compactness can be viewed as part of the {\em theoretical orientation}.
Proving these theorems on the expense of using the meaning function to teach reading and writing $\SOL$ and $\FOL$
formulas amounts to depriving the students of the practical knowledge they need later on.

\section{Postscript}
Since the publication of \cite{conf-lpar-Makowsky07,makowsky2008a-hilbert}
I had many discussions on these topics. I also received feedback on the current paper,
from the referees, and otherwise. 
Many raised  objections to my views, however, often these were
the result of misunderstandings.

\paragraph*{Sets and Logic for Undergraduates}
In \cite{conf-lpar-Makowsky07,makowsky2008a-hilbert} I discussed how to teach logic to CS-students in general.
I had a Logic Course in mind similar to the traditional ``Cours d'analyse'' of the Grandes \'Ecoles in France
still a tradition before 1968, something like Shoenfield's book, \cite{bk:shoe}, but for CS students.
I was told by my French colleagues that this was too elitist, that even these ``Cours d'analyse'' were not anymore
what they used to be.

In this paper I ask the question whether one should have compulsory courses covering
the basics of modeling artifacts as sets and the basics of logic already in the three year program of
computer science. I also assumed that these students were likely to take later courses such as
{\em Introduction to Database}, {\em Automata and Formal Languages},  and some courses which confront 
them with the basics of {\em Programming Languages} and {\em Hardware and Software Verification}.
I assumed further that a one semester (12 weeks) course of three frontal lectures and two tutorial sessions
per week was available.

I emphatically defend teaching Sets and Logic to the undergraduates.
But facing the threat that by teaching too much material the department colleagues 
will vote Logic out of the undergraduate program I took great care to balance the topics
with a  view to other  courses in the undergraduate program.
I tailored the syllabus of the proposed course fitting my assumptions above.
As the three year (or even four year) CS programs tend to be very loaded,
I suggested to skip many topics dear to logicians or practitioners of logic.
I presented my views on the basis of uniting 
the basics of modeling artifacts as sets and the basics of logic 
in one course. There is no need for that, but there are distinct advantages in doing so.
Spreading the material over two or more courses, which is frequently done, only dilutes
the material, leads to unnecessary repetitions, and creates problems  of coordination
among the teachers of these courses.

\paragraph*{Where is All the Logic Gone?}
It was pointed out to me that topics I suggested to skip, such as the proof of the completeness theorem,
are needed in courses dealing with {\em  Artificial Intelligence}, especially for description logics.
True! Similar objections can be made also in favor of modal logic, temporal logic, proof theory,
finite model theory , etc., etc. 
Well, all these nice and dear topics belong, from my point of view, to specialized graduate programs.
The emergence of specialized graduate programs such as Logic in CS, Logic and Computation, and the like,
especially in Europe (Vienna, Dresden, Trento) lends strong support to such a view. 

\paragraph*{Textbooks}
My dear readers suggested that my list of books recommended was incomplete. Yes, indeed, it is.
However, none of the books they suggested to include was written after 1990, and none of these books
went beyond the traditional narrative.


\newpage
\thispagestyle{empty}
{\ }

\end{document}